\documentclass[reprint,prl,twocolumn,superscriptaddress,showpacs,nofootinbib,floatfix,preprintnumbers]{revtex4-1}
\usepackage{graphicx, bm}
\newcommand{\lsim}{\mbox{\raisebox{-.6ex}{~$\stackrel{<}{\sim}$~}}}
\newcommand{\gsim}{\mbox{\raisebox{-.6ex}{~$\stackrel{>}{\sim}$~}}}

\begin{document}
\preprint{MAN/HEP/2013/08, IPPP/13/26, DCPT/13/52}
\title{Constraining Neutrino Mass from Neutrinoless Double Beta Decay}

\author{P. S. Bhupal Dev}
\affiliation{Consortium for Fundamental Physics, School of Physics and Astronomy, University of Manchester, Manchester M13 9PL, United Kingdom}

\author{Srubabati Goswami}
\affiliation{Physical Research Laboratory, Navrangpura, Ahmedabad  380009, India}

\author{Manimala Mitra}
\affiliation{Institute for Particle Physics Phenomenology, 
Department of Physics, Durham University, Durham DH1 3LE, United Kingdom}

\author{Werner Rodejohann}
\affiliation{Max-Planck-Institut f\"{u}r Kernphysik,
Saupfercheckweg 1, 69117 Heidelberg, Germany}

\begin{abstract}
We re-analyze the compatibility of the claimed observation of 
neutrinoless double beta decay 
($0\nu\beta\beta$) in $^{76}$Ge with the new limits on the 
half-life of $^{136}$Xe from EXO-200
and KamLAND-Zen. Including recent calculations of the 
nuclear matrix elements (NMEs), we show that while the claim in $^{76}$Ge is still compatible with the individual limits from $^{136}$Xe for a few NME calculations, it is inconsistent with the KamLAND-Zen+EXO-200 combined  limit  for all but one NME. 
 After imposing the most stringent upper limit on the 
sum of  light neutrino masses  from Planck,  
we find that the canonical 
light neutrino contribution cannot satisfy the claimed $0\nu\beta\beta$ 
signature or saturate the current limit, irrespective of the NME 
uncertainties. However,  inclusion of the heavy neutrino contributions,
arising naturally in TeV-scale Left-Right symmetric models, can 
saturate the current limit of $0\nu\beta\beta$. In a type-II seesaw framework, this  imposes a lower limit on the 
lightest neutrino mass. 
Depending on the mass hierarchy, we obtain this limit to be in the range of 0.07 - 4 meV for a typical choice of the right-handed (RH) gauge boson and RH neutrino masses  
relevant for their collider searches.  
Using the $0\nu\beta\beta$ bounds, we also derive correlated constraints in the RH sector, complimentary to those from the 
LHC. 
 \end{abstract}
\maketitle
\textit {\textbf {Introduction}} -- 
The discovery of neutrino oscillations, and hence, non-zero neutrino masses and mixing  
implies physics beyond the Standard Model (SM).  
Some of the unresolved issues 
are (i) whether neutrinos are Majorana or Dirac particles, 
(ii) their absolute mass scale, and (iii) their mass hierarchy.  
Neutrinoless double beta decay ($0\nu\beta\beta$)~\cite{review}, 
if observed, 
would imply lepton number violation (LNV) and  Majorana nature of
neutrinos~\cite{Schechter:1981bd}, and 
could possibly shed light on  
the other issues.

Experimental studies of the $0\nu\beta\beta$ process: $(A,Z)\to (A,Z+2)+2e^-$ 
have been conducted on several nuclei, and to date, there has been only one claimed 
observation in $^{76}$Ge with half-life $T_{1/2}^{0\nu}(^{76}\rm{Ge})=2.23^{+0.44}_{-0.31}\times 10^{25}$ yr 
at 68\% CL~\cite{Klapdor:2006ff}. 
Several ongoing experiments have 
design
sensitivities to test this claim. 
Recently, the KamLAND-Zen (KLZ) experiment using $^{136}{\rm Xe}$ obtained  
the  limit $T_{1/2}^{0\nu}(^{136}\rm{Xe})>1.9\times 10^{25}$ yr at 90\% CL~\cite{Gando:2012zm}. After combining with the EXO-200 (EXO) results, $T_{1/2}^{0\nu}(^{136}\rm{Xe})>1.6\times 10^{25}$ yr~\cite{Auger:2012ar}, they derived the limit  
$T_{1/2}^{0\nu}(^{136}\rm{Xe})>3.4\times 10^{25}$ yr at 
90\% CL~\cite{Gando:2012zm}, and disfavored the claim in~\cite{Klapdor:2006ff} at $>97.5\%$ CL,  
using recent 
calculations of the nuclear matrix elements (NMEs).

On the other hand, the 
Planck results in conjunction with 
other cosmological data have put a stringent upper 
limit on the sum of light neutrino masses: $\sum m_\nu <0.23$ eV at 95\% CL~\cite{Ade:2013lta},  
which rules out most of the quasi-degenerate region of the light neutrino mass spectrum. 
This has important consequences for the canonical interpretation of 
$0\nu\beta\beta$  via light neutrino exchange 
\cite{fogli}.

In this paper we study the implications of these recent results on 
various aspects of the 
$0\nu\beta\beta$ phenomenology, namely, we   
(i) re-analyze the compatibility of the  
KamLAND-Zen and EXO-200 limits with the claimed observation~\cite{Klapdor:2006ff}, 
including 
the uncertainties due to
several updated
NME calculations; 
(ii) quantify whether the standard light neutrino prediction for 
$0\nu \beta\beta$ can satisfy  the claimed observation or
saturate the current limit, while being consistent with the stringent 
neutrino mass constraints from cosmology;  and
(iii) investigate whether a
heavy neutrino contribution naturally arising in low scale 
Left-Right symmetric models (LRSM), 
accessible at the LHC, can saturate the $0\nu\beta\beta$ limit.

{\textit{\textbf{Light Neutrino Contribution}}}--
For  $0\nu\beta\beta$ mediated by the light Majorana neutrinos, 
the half-life is given by 
\begin{eqnarray}
\frac{1}{T_{1/2}^{0\nu}} 
= G_{0\nu}|{\cal M}_\nu|^2\left|\frac{m^\nu_{ee}}{m_e}\right|^2\, ,
\label{half1}
\end{eqnarray}
where $G_{0\nu}$, ${\cal M}_{\nu}$ and $m_e$ 
are the  
the phase space factor, the NME, 
and the electron mass respectively.  
Here $m^\nu_{ee} = \sum_i U_{ei}^2 m_i$
is the effective mass, where $U$ is the PMNS mixing matrix diagonalizing the light neutrino mass matrix with eigenvalues $m_i~(i=1,2,3)$. Using the standard parametrization for $U$,  
we obtain (with $c_{ij}\equiv \cos\theta_{ij},s_{ij}\equiv \sin\theta_{ij}$)
\begin{eqnarray}
m^\nu_{ee} = m_1c_{12}^2c_{13}^2+m_2s_{12}^2c_{13}^2e^{2i\alpha_2}+m_3s_{13}^2e^{2i\alpha_3} \, .
\label{mnuee}
\end{eqnarray}  
To test  the compatibility between the claim in 
~\cite{Klapdor:2006ff} and 
the null results in~\cite{Gando:2012zm, Auger:2012ar}, it is useful to
study the correlation between their half-lives (see also~\cite{faess}) using Eq.~(\ref{half1}): 
\begin{eqnarray}
T_{1/2}^{0\nu}(^{136}{\rm Xe}) 
&=& \left(3.61^{+1.18}_{-0.83}\times 10^{24}~{\rm yr}\right)\left|\frac{{\cal M}_{0\nu}(^{76}{\rm Ge})}{{\cal M}_{0\nu}(^{136}{\rm Xe})}\right|^2
\label{gexe}
\end{eqnarray}
where we have used the recently re-evaluated 
phase space factors~\cite{Kotila:2012zza} 
for the axial-vector coupling constant $g_A=1.25$.
We take the claimed value for $T_{1/2}^{0\nu}(^{76}{\rm Ge})$~\cite{Klapdor:2006ff} at 90\% CL  (assuming Gaussian errors). An experimental limit on $T_{1/2}^{0\nu}(^{136}{\rm Xe})$ larger than the predicted value 
from Eq.~(\ref{gexe}) will rule out the positive claim of~\cite{Klapdor:2006ff}.
Using various updated 
NME calculations~\cite{Rodriguez:2010mn, Menendez:2008jp, Barea:2013bz, Suhonen:2010zzc, Meroni:2012qf, 
Simkovic:2013qiy, Mustonen:2013zu}, 
we show in Table~\ref{tab1} the predicted range of $T_{1/2}^{0\nu}(^{136}{\rm Xe})$ at 90\% CL.  Note that for a given NME method, when different versions of the results are available, we only quote the extreme (smallest and largest) values to show the allowed ranges.  
We find  that it is still compatible with the individual 
limits from KLZ and EXO for some of the NMEs calculated by QRPA method~\cite{Meroni:2012qf, 
Simkovic:2013qiy, Mustonen:2013zu}, but inconsistent with their 
combined limit 
in~\cite{Gando:2012zm} for all of the NME values, 
except the one given in \cite{Mustonen:2013zu}. 
The reason is the very small NME for $^{136}$Xe in \cite{Mustonen:2013zu}, 
which can be 
attributed to the 
differences in pairing structure in the neutron mean fields, thus leading to a small overlap in the initial and final mean fields.

\begin{table}[h!]
\begin{center}
\begin{tabular}{c|c|c|c}\hline\hline
\multicolumn{3}{c|}{NME} & $T_{1/2}^{0\nu}(^{136}{\rm Xe})$\\ \cline{1-3}
Method & ${\cal M}_{0\nu}(^{76}{\rm Ge})$ & ${\cal M}_{0\nu}(^{136}{\rm Xe})$ & [$10^{25}$ yr]\\ \hline
EDF(U)~\cite{Rodriguez:2010mn} & 4.60 & 4.20 & 0.33 - 0.57 \\ 
ISM(U)~\cite{Menendez:2008jp} & 2.81 & 2.19 &  0.46 - 0.79 \\ 
IBM-2~\cite{Barea:2013bz} & 5.42 & 3.33 & 0.74 - 1.27 \\ 
pnQRPA(U)~\cite{Suhonen:2010zzc} & 5.18 & 3.16 & 0.75 - 1.29 \\
SRQRPA-B~\cite{Meroni:2012qf} & 5.82 & 3.36 & 0.84 - 1.44 \\ 
SRQRPA-A~\cite{Meroni:2012qf} & 4.75 & 2.29 & 1.20 - 2.06 \\ 
QRPA-B~\cite{Simkovic:2013qiy} & 5.57 & 2.46 & 1.43 - 2.46 \\ 
QRPA-A~\cite{Simkovic:2013qiy} & 5.16 & 2.18 & 1.56 - 2.69 \\ 
SkM-HFB-QRPA~\cite{Mustonen:2013zu} & 5.09 & 1.89 & 2.02 - 3.47 \\ 
\hline\hline
\end{tabular}
\end{center}
\caption{Predictions for $T_{1/2}^{0\nu}(^{136}{\rm Xe})$ at $ 90\% $ CL corresponding to the 
claimed $0\nu\beta\beta$ observation in $^{76}$Ge~\cite{Klapdor:2006ff} for 
the latest results of different NME calculations~\cite{Rodriguez:2010mn, Menendez:2008jp, Barea:2013bz, Suhonen:2010zzc, Meroni:2012qf, Simkovic:2013qiy, Mustonen:2013zu}.  
}
\label{tab1}
\end{table}

\begin{figure*}[ht]
\centering
\includegraphics[width=6cm]{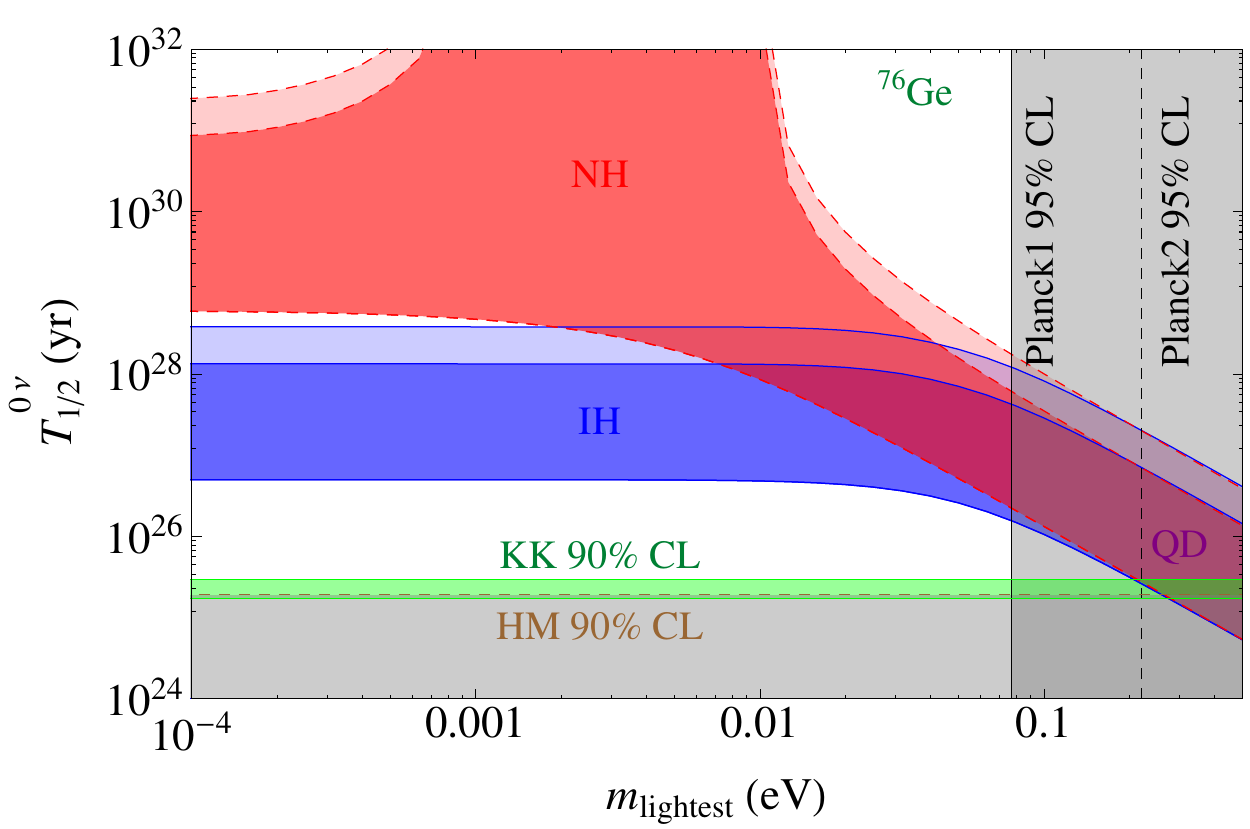}
\hspace{0.5cm}
\includegraphics[width=6cm]{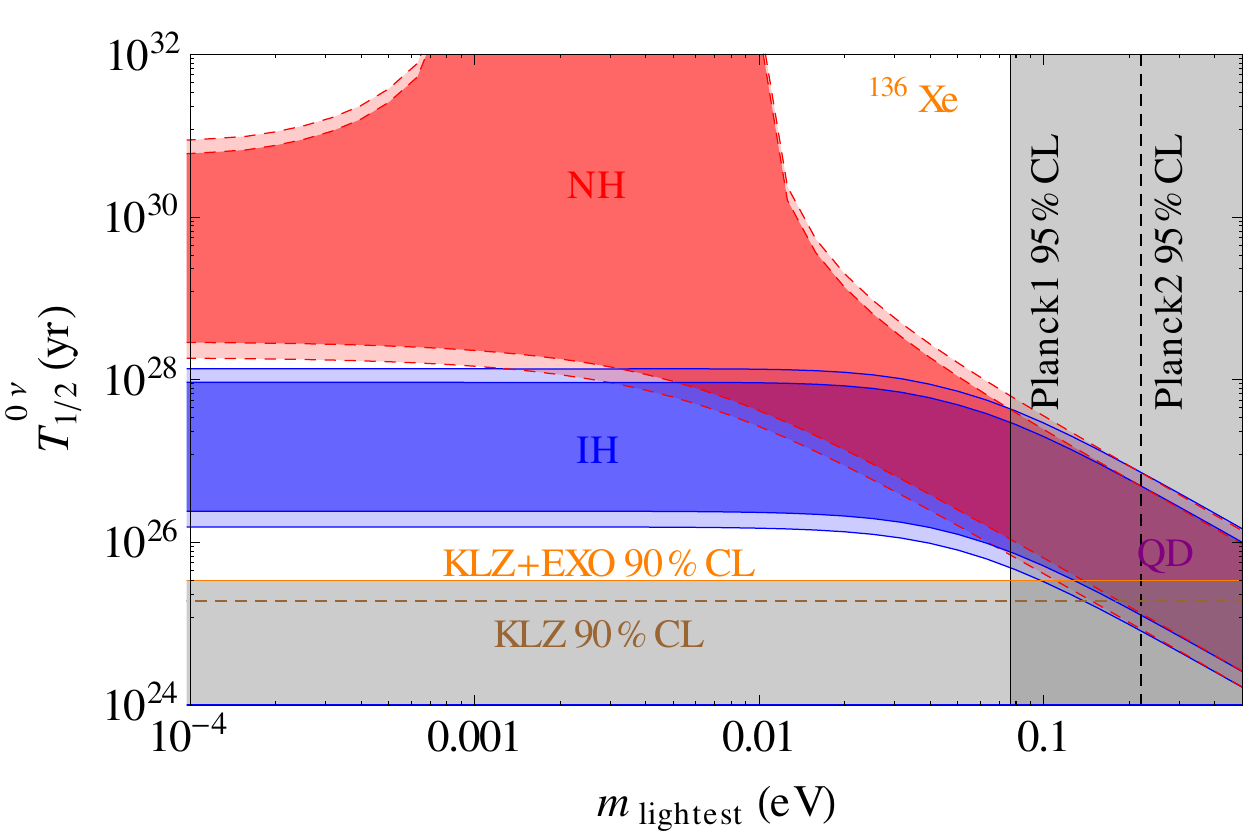}
\caption{The predicted half-life  of
$0\nu\beta\beta$ in $^{76}$Ge (left) and $^{136}$Xe (right) due to light  
neutrino exchange. The light shaded regions include the uncertainties due to all the NMEs listed in Table~\ref{tab1}, whereas the dark shaded regions correspond to the NMEs in~\cite{Meroni:2012qf}. The grey regions are excluded from 
$0\nu\beta\beta$ and Planck results (see text for details). 
}
\label{fig1}
\end{figure*}

For comparison of the experimental results with the canonical light neutrino contribution given by Eq.~(\ref{half1}) including all the NME uncertainties, it is better to consider the individual half-lives of 
different isotopes (instead of the effective mass which is theoretically 
independent of the NME uncertainties). Hence, we show in Fig.~\ref{fig1} the predicted half-lives for $^{76}$Ge and $^{136}$Xe  
as a function of the lightest neutrino mass for  normal and inverted mass orderings, including the hierarchical 
and quasi-degenerate (QD) regimes.    
We have varied  the oscillation parameters in their $3\sigma$  
 range~\cite{global}, the $C\!P$ phases from 0 to $\pi$, and 
included  the  NME uncertainties from  
Table~\ref{tab1}  (light shaded regions). 
Note that 
the predicted regions of half-life for normal hierarchy (NH) and inverted hierarchy (IH) almost overlap due to the  
NME uncertainties. 
However, for a given set of NMEs 
(e.g., those of~\cite{Meroni:2012qf} taken here for illustration),
 we recover the standard picture with the two (dark shaded) regions  well-separated. The green (solid) horizontal lines in the left panel of Fig.~\ref{fig1} correspond to 
the 90\% CL claim value of \cite{Klapdor:2006ff} (KK), and the brown (dashed) horizontal line for the lower limit set by the Heidelberg-Moscow collaboration~\cite{hdlbrg} (HM). The orange (solid) and brown (dashed) horizontal lines in the right panel represent the  90\% CL lower limits for $^{136}$Xe from KLZ and 
combined KLZ+EXO~\cite{Gando:2012zm} respectively.
The solid vertical line shows the 95\% CL limit,  
$\sum m_\nu<0.23$ eV (Planck1), derived from the Planck+WMAP low-multipole polarization+high resolution CMB+baryon acoustic oscillation (BAO) data and assuming a standard $\Lambda$CDM model of cosmology, whereas the dashed vertical line shows the 
limit without the BAO data set:  $\sum m_\nu<0.66$ eV (Planck2)~\cite{Ade:2013lta}.  Note 
that although the cosmological bound on the sum of neutrino masses depends strongly on the choice of data sets, it is currently stronger than the direct experimental bound coming from Tritium $\beta$ decay experiment: $m_{\nu_e} \lsim 2$ eV~\cite{tritium}. 

The current constraints on $0\nu\beta\beta$ (including the claim) 
can be saturated by the canonical contribution only in the QD regime 
with $m_1\simeq m_2\simeq m_3\equiv m_0 \gsim 0.1$ eV. As it is evident from Fig.~\ref{fig1}, this possibility is excluded, 
regardless of the NME uncertainties, if we take the most stringent 
upper limit from cosmology which for QD neutrinos gives $m_0<0.077$ eV. 
For other cosmological data sets, only a very narrow allowed mass window 
remains.

{\textit {\textbf{Heavy Neutrino Contribution}}}-- 
The heavy right-handed (RH) neutrinos, introduced in  
the type-I seesaw \cite{type1} models, 
if sufficiently light $(\leq 10 \, \rm{TeV})$, 
can give a significant contribution 
to $ 0 \nu\beta\beta$~\cite{Ibarra:2010xw} 
provided their mixing with the active neutrinos is sizeable. 
However, this requires fine-tuning and/or cancellation 
\cite{Pilaftsis:1991ug}. A more natural way to obtain appreciable heavy neutrino contributions to the 
$0\nu\beta\beta$ amplitude arises in the TeV scale
LRSM~\cite{LRSM} via RH  
currents~\cite{Tello:2010, Nemevsek:2011aa}. 
Such models 
also lead to other high and low-energy phenomena 
and could for instance be directly probed at  
the LHC through the same-sign dilepton signal~\cite{KS}.

The LRSM includes heavy neutrinos as part of the $SU(2)_R$ doublet and 
restores parity at high energies~\cite{LRSM}. This naturally 
leads to small 
neutrino masses through either type-I seesaw
via the RH neutrinos~\cite{type1} or type-II seesaw via $SU(2)$ triplet scalars~\cite{type2} or both~\cite{mixed}. The corresponding Lagrangian is given by 
\begin{eqnarray}
{\cal L}_{Y} &=& f_\nu \bar{L}_L\Phi L_R+ \widetilde{f}_\nu \bar{L}_L\widetilde{\Phi} L_R
+f_L L_L^{\sf T}Ci\sigma_2\Delta_L L_L\nonumber \\ 
&& +f_R L_R^{\sf T}Ci\sigma_2\Delta_R L_R
+{\rm h.c.} 
\label{lag}
\end{eqnarray}
Here $C$ is the charge conjugation operator and $\sigma_2$ the second Pauli 
Matrix, $L_{L(R)}$ denotes the 
lepton doublet, $\Phi$ the SM Higgs doublet, $\widetilde{\Phi}=\sigma_2\Phi^*\sigma_2$, and $\Delta_{L(R)}$ the 
scalar triplet belonging to
$SU(2)_{L(R)}$. The light neutrino mass matrix in the 
seesaw approximation is $M_\nu \simeq m_L-m_D^{\sf T}M_R^{-1}m_D$,
where $m_D=f_\nu v$, $m_L=f_L v_L$,  $M_R=f_R v_R$, and $v$, $v_{L(R)}$ are the vacuum expectation values of doublet and triplet Higgs fields:
$\langle \Phi\rangle=v,~\langle \Delta_{L(R)}\rangle=v_{L(R)}$. 
The heavy neutrino masses $\sim M_R$ are
related to the RH gauge boson mass $M_{W_R}=gv_R$. 

There are several diagrams leading to double beta decay in LRSM (see~\cite{review} and references therein).  
In this work we  consider  
the appealing case of 
type-II dominance~\cite{Tello:2010}. 
Also, the scalar triplet contribution is expected to be small due to
constraints from lepton flavor violation, 
which typically require $M_N/M_\Delta\lsim 0.1$~\cite{Tello:2010}. 
Hence, we focus only on the 
diagram with purely RH currents, mediated by the heavy neutrinos
which adds coherently to the purely left-handed light neutrino contribution 
discussed earlier: 
\begin{eqnarray}
\frac{1}{T_{1/2}^{0\nu}} = G_{0\nu}|{\cal M}_\nu|^2\left|\frac{m_{ee}^{(\nu+N)}}{m_e}\right|^2 , 
\label{half2}
\end{eqnarray}
where $\left|m_{ee}^{(\nu+N)}\right|^2=|m_{ee}^\nu|^2+|m_{ee}^N|^2$, with $m_{ee}^\nu$ given by Eq.~(\ref{mnuee}) and $m_{ee}^N$ is the heavy neutrino effective mass: 
\begin{eqnarray}
m_{ee}^N = \langle p^2\rangle\frac{M_{W_L}^4}{M_{W_R}^4}\sum_j\frac{V_{ej}^2}{M_j}  \, .
\label{mNee}
\end{eqnarray} 
Here $\langle p^2\rangle =-m_em_p{{\cal M}_N}/{{\cal M}_\nu}$
denotes the virtuality of the exchanged neutrino,    
$m_p$ is the mass of the proton and ${\cal M}_N$ is the NME corresponding to the RH neutrino exchange. Note that Eq.~(\ref{mNee}) is valid only in the heavy neutrino limit: $M_j^2\gg |\langle p^2\rangle|$ which is assumed hereafter. Using the values for ${\cal M}_{\nu}$ and ${\cal M}_N$ from~\cite{Meroni:2012qf}, we get  
$\langle p^2\rangle =  -$(157 - 185 MeV)$^2$ for $^{136}$Xe and $-$(153 - 184 MeV)$^2$ for $^{76}$Ge.  
The unitary matrix $V$ in Eq.~(\ref{mNee}) diagonalizes
$M_R$ with mass eigenvalues $M_j$. 
We further assume the discrete LR symmetry to be 
parity, under which $f_L=f_R$ and $U=V$. Our conclusions remain unchanged for 
the other possibility viz.\ charge conjugation: $f_L=f_R^*$ and $U=V^*$.
\begin{figure*}[htb]
\centering
\includegraphics[width=6cm]{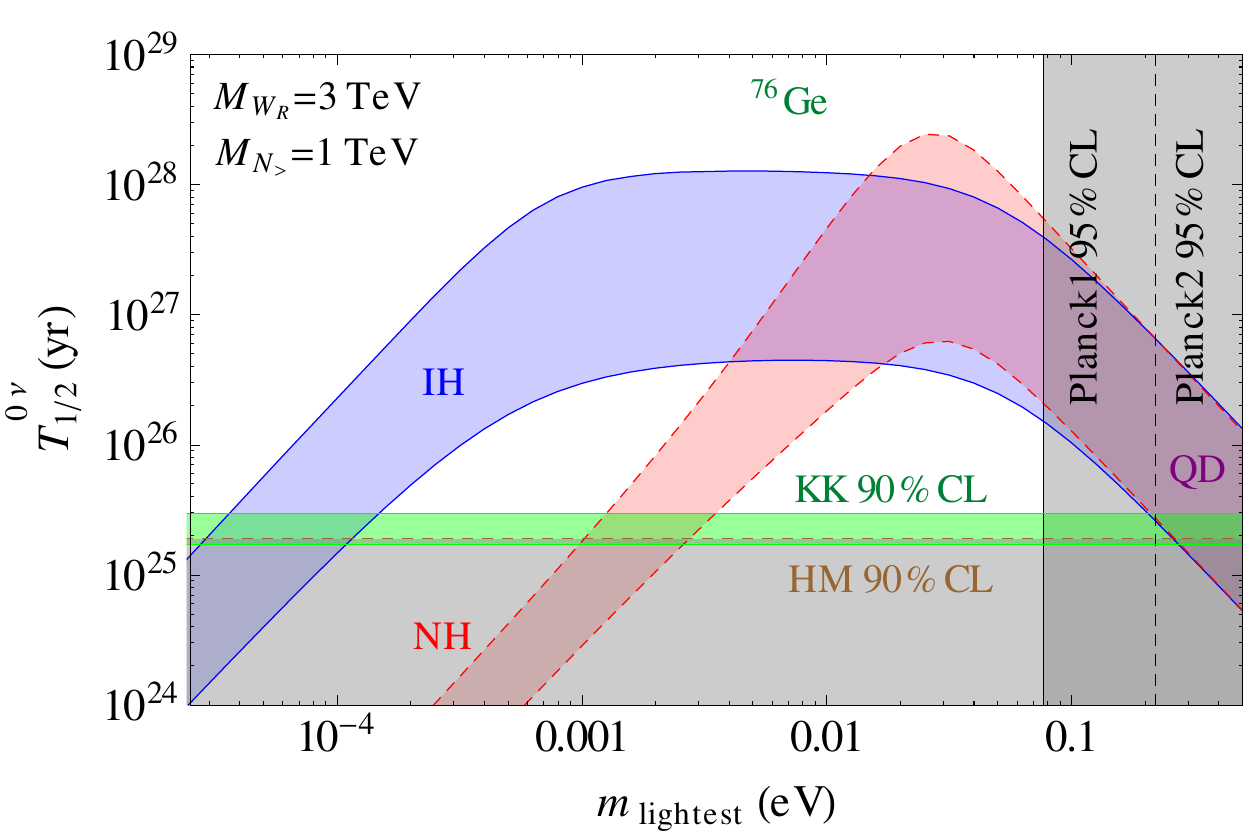}
\hspace{0.5cm}
\includegraphics[width=6cm]{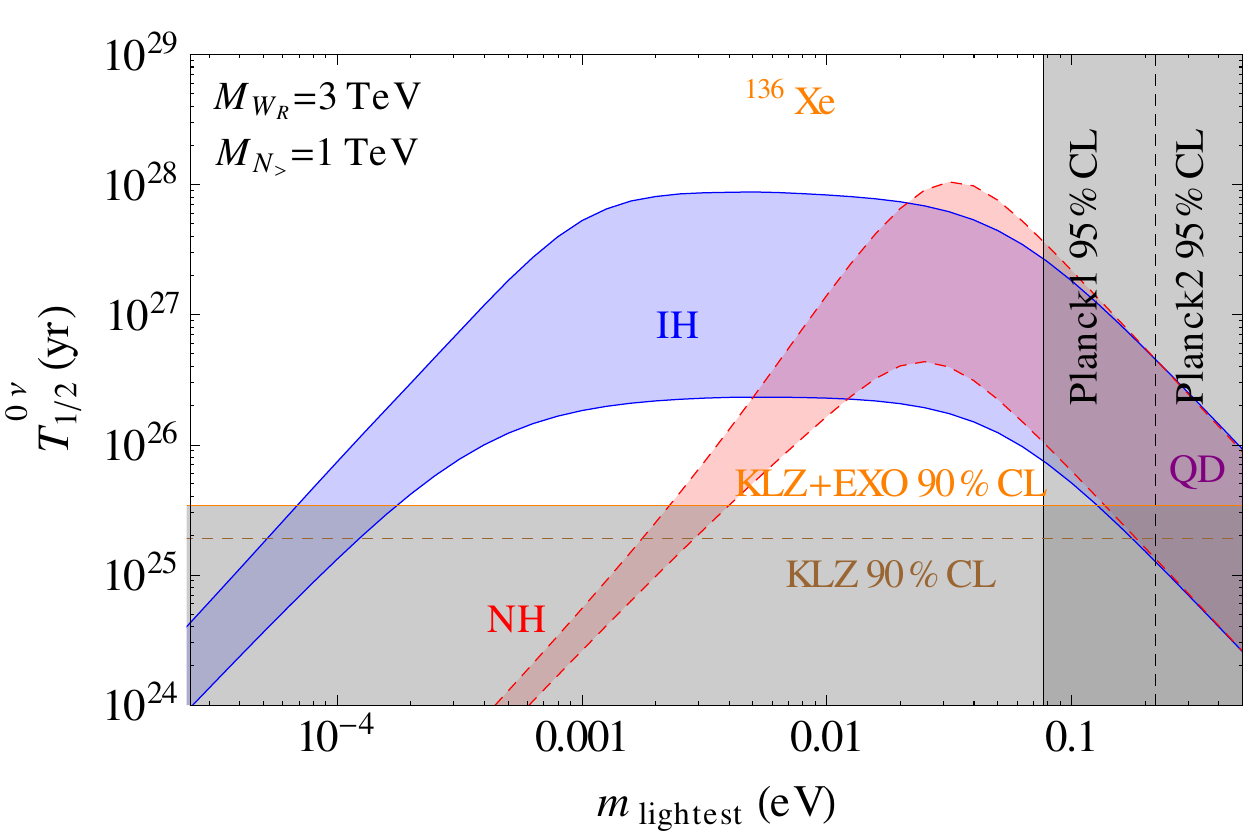}
\caption{
The light+heavy neutrino contribution to the $0\nu\beta\beta$ half-life of $^{76}\rm{Ge}$ (left) and $^{136}\rm{Xe}$ (right) 
for both NH and IH, and with type-II seesaw dominance. Here 
$(M_{W_R},M_{N_>})=(3,1)$ TeV. 
The vertical and horizontal lines are same as in Fig.~\ref{fig1}. }
\label{fig2}
\end{figure*}

In the type-II limit, 
 $M_\nu \simeq m_L=(v_L/v_R)M_R$ and 
$m_i\propto M_i$. Hence, for the normal ordering we have $M_1<M_2\ll M_3$  as well, and the RH neutrino masses can be expressed in terms of the heaviest one as $M_1/M_3=m_1/m_3,~M_2/M_3=m_2/m_3$. 
Then 
\begin{eqnarray}
m_{ee}^N|_{\rm nor}=\frac{C_N}{M_3}\left(\frac{m_3}{m_1}c_{12}^2c_{13}^2+\frac{m_3}{m_2}s_{12}^2c_{13}^2e^{2i\alpha_2}+s_{13}^2e^{2i\alpha_3}\right),\nonumber
\end{eqnarray}
where $C_N = {\langle p^2\rangle}M_{W_L}^4/M_{W_R}^4$. 
For inverted ordering, $M_2$ will be the largest, 
and hence 
\begin{eqnarray}
m_{ee}^N|_{\rm inv}= \frac{C_N}{M_2}\left(\frac{m_2}{m_1}c_{12}^2c_{13}^2+s_{12}^2c_{13}^2e^{2i\alpha_2}+\frac{m_2}{m_3}s_{13}^2e^{2i\alpha_3}\right).\nonumber
\end{eqnarray} 
In Fig.~\ref{fig2}, we show the half-life predictions for $^{76}$Ge and $^{136}$Xe  
using Eq.~(\ref{half2}), and including the light and heavy neutrino NME ranges given in~\cite{Meroni:2012qf} (corresponding to  $g_A=1.25$).  Here we have chosen  
$M_{W_R}$ = 3 TeV and the heaviest neutrino mass, $M_{N_>}$= 1 TeV, keeping in mind the current LHC exclusion limits~\cite{atlas-LR} and its future accessible range. Note that for this choice of $M_{N_>}$, and for the range of the lightest neutrino mass shown in Fig.~\ref{fig2}, the lightest RH neutrino mass is 
$M_{N_<}>490$ MeV, which justifies the validity of  Eq.~(\ref{mNee}). Several important conclusions can be drawn from this illustrative plot: 
(i) the purely RH  
contribution via exchange of 
heavy neutrinos, when added to the standard light neutrino contribution, 
can  saturate the current experimental limit (or satisfy the claim) even for 
hierarchical neutrinos;  
(ii) for the heavy neutrino contribution 
saturating the bound on $T_{1/2}^{0\nu}$, there exists an absolute {\it lower} bound on the lightest neutrino mass  both for orderings: 
(2 - 4) meV for NH and  (0.07 - 0.2) meV for IH. The range is due to the 
combined effect of the NME uncertainties and the $3\sigma$ range of the oscillation parameters used here. Needless to mention, the lower bound will become stronger with improved experimental bounds on $0\nu\beta\beta$ in future.    
(iii) the KK claim can be 
reached for 
the lightest neutrino mass in the range of  (1 - 3) meV for NH and 
(0.03 - 0.1) meV for IH. 
These values are well within  the most stringent Planck limit 
of 77 meV;  
(iv) for the heavy neutrino contribution, the compatibility between the KK claim and KLZ+EXO bound  can be examined using  Eq.~(\ref{gexe}), with the NMEs for light neutrinos replaced by those for heavy neutrinos~\cite{Meroni:2012qf}. It 
predicts the half-life for $^{136}$Xe in the range $(0.56-2.74) \times 10^{25}\, \rm{yr}$ 
 at 90\% CL,  for all the corresponding NMEs in \cite{Meroni:2012qf}.  Thus 
in this case also,  the KK claim is compatible with the individual KLZ and EXO bounds, but inconsistent  with their combined limit.
Similar conclusion holds for the light+heavy neutrino  contribution, since the KK claim can be saturated while being consistent with cosmology only by a dominant heavy neutrino contribution; 
(v) the lower bound is sensitive to the 
RH neutrino and gauge boson masses.  For a given $W_R$ mass, the lower bound on $m_{\rm lightest}$ is weakened by increasing the RH neutrino mass $M_{N>}$, and the bound tightens for lower $M_{N>}$ (as long as we are in the heavy neutrino regime so that Eq.~(\ref{mNee}) is valid; otherwise, no lower limit on $m_{\rm lightest}$ can be derived).  
The trend is similar if we vary the $W_R$ mass, but more pronounced due to the $M_{W_R}^{-4}$ dependence in Eq.~(\ref{mNee}).

\begin{figure*}[ht]
\centering
\includegraphics[width=6cm]{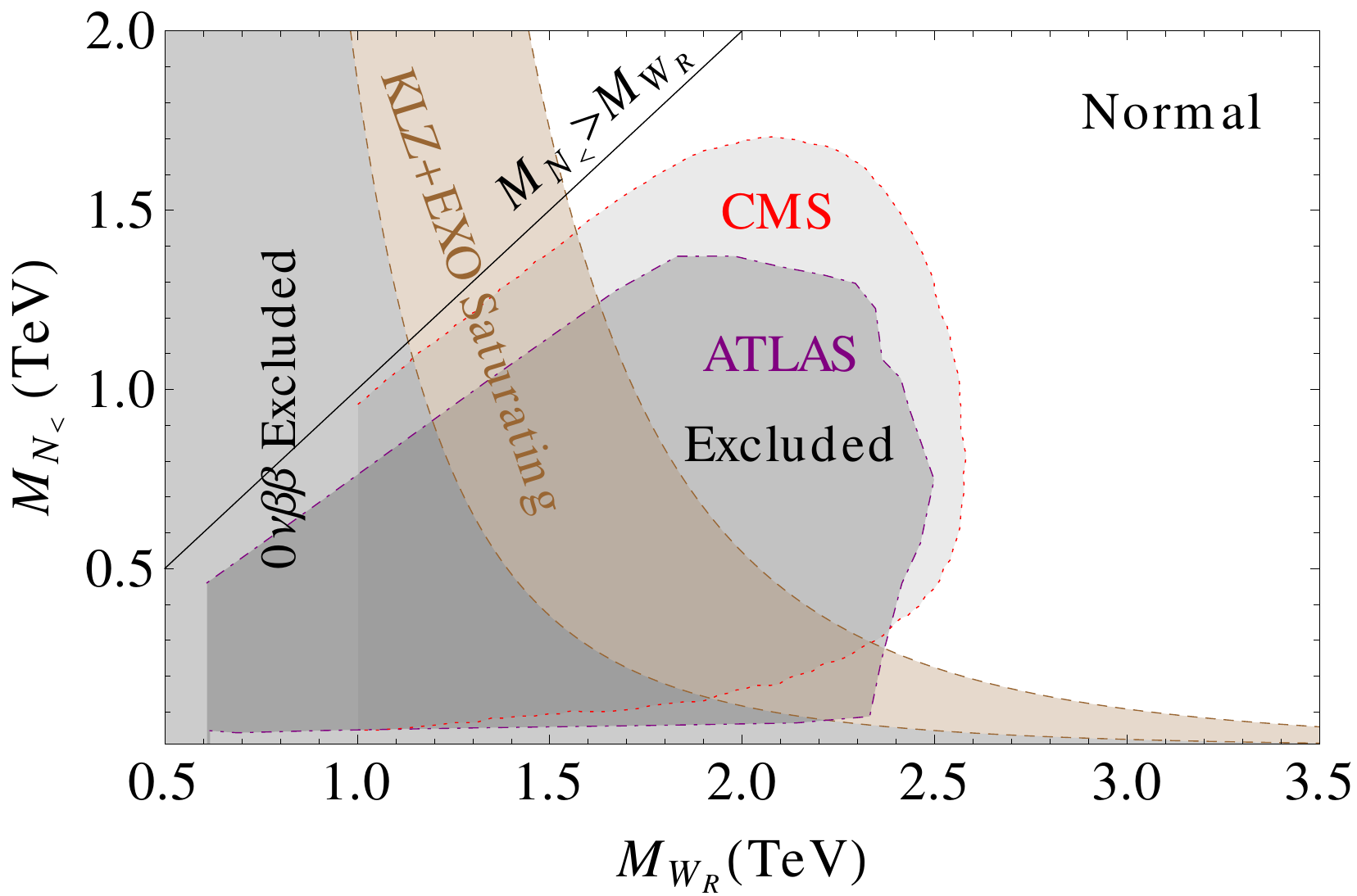}
\hspace{0.5cm}
\includegraphics[width=6cm]{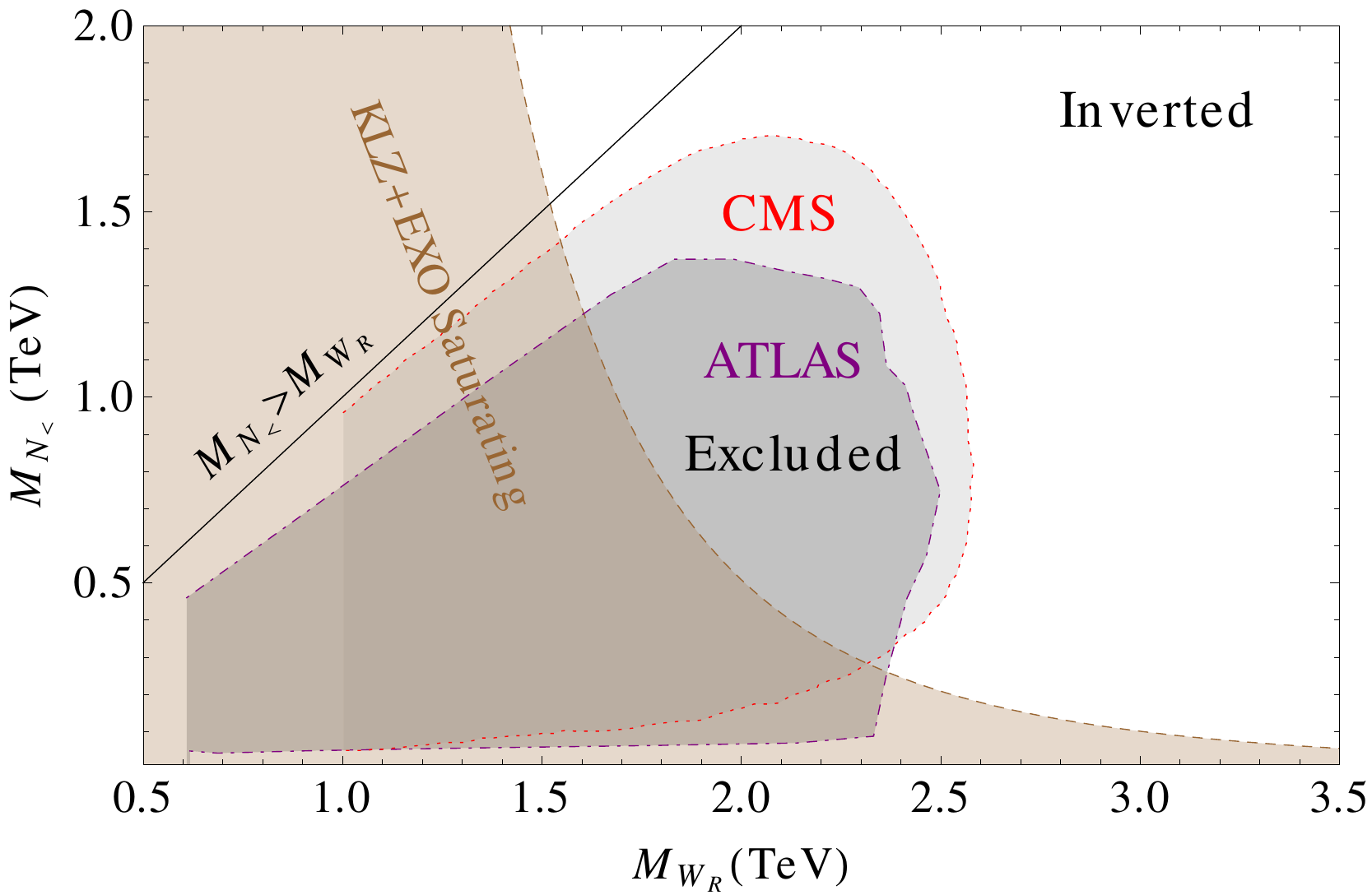}
\caption{The $0\nu\beta\beta$ constraints in the $M_{W_R}$-$M_{N_<}$ plane, along with the direct search limits from CMS and ATLAS. The brown (dashed) 
region saturates the KLZ+EXO combined limit, and the grey (white) 
region is excluded (allowed).
}
\label{fig4}
\end{figure*}

{\textit {\textbf {Complementarity with the LHC results}}} --  $0\nu\beta\beta$ provides a 
complementary probe to collider searches for LNV. 
The correlation between the heavy gauge boson mass and the lightest RH neutrino mass for a TeV-scale LRSM is shown in Fig.~\ref{fig4}  for both mass orderings. In the brown (dashed) shaded region, the half-life in Eq.~(\ref{half2}) saturates the combined limit from KLZ+EXO~\cite{Gando:2012zm}, whereas the region to 
its left (right) is excluded (allowed) by this limit. 
The width of the brown region is due to the  
variation of 
the oscillation parameters in their 3$\sigma$
range~\cite{global} and the lightest neutrino 
mass up to the most stringent upper limit from Planck.
We have considered the  NMEs for $^{136}$Xe  
corresponding to light and heavy neutrino exchange
\cite{Meroni:2012qf} which yield the smallest $|\langle p^2\rangle|$, 
and hence, the strongest limit in Fig.~\ref{fig4}. 
The current LHC exclusion regions~\cite{atlas-LR} are also shown for 
comparison (see also \cite{Das:2012ii} for detail discussion on collider searches). 
We find that (i) for the normal ordering, a  
part of the parameter space not accessible at the LHC can be constrained (or 
probed in case of an observation) through $0\nu\beta\beta$, and 
(ii) for the inverted ordering, it is not possible to exclude any parameter space 
in the $M_{W_R}-M_{N_<}$ plane from $0\nu\beta\beta$
due to cancellations in $m_{ee}^N$.

{\textit {\textbf {Conclusion}}}-- In summary, (i) the positive claim of $0\nu\beta\beta$ in $^{76}$Ge is still compatible with the individual $^{136}$Xe limits from
EXO-200 and KamLAND-Zen due to NME uncertainties, whereas the combined limit 
excludes this 
for all but one 
NME calculations; 
(ii) the most stringent limit on 
$\sum m_\nu$
from Planck,
in conjunction 
with the KamLAND-Zen+EXO-200  bound, 
excludes the possibility of saturating the 
limit for $^{136}$Xe 
or the claim in $^{76}$Ge solely by the canonical light neutrino contribution; 
(iii) the additional heavy
neutrino contribution to $0\nu\beta\beta$ via purely RH currents 
in the TeV-scale minimal Left-Right extension of the SM
can saturate the current experimental bound. For type-II seesaw dominance,  it sets
a lower limit on the lightest neutrino mass;  (iv) we show for normal mass ordering, $0\nu \beta \beta$ puts additional 
constraints  in the RH gauge boson and heavy neutrino mass plane, 
complementary to those from LHC. 

{\textit{\textbf{Acknowledgments}}}-- 
PSBD and MM thank C.\ Boehm, F.\ Deppisch, S.\ Pascoli, and F.\ Vissani 
 for  discussions. This work is supported by 
STFC 
grant ST/J000418/1 (PSBD), DFG-DST project RO 2516/4-1 (SG and WR), 
European ITN project
FP7-PEOPLE-2011-ITN, PITN-GA-2011-289442-INVISIBLES (MM) 
and the Max Planck Society 
project MANITOP through the Strategic Innovation Fund (WR).
PSBD, MM, and SG thank the organizers of BLV2013 and MPIK, Heidelberg for hospitality. 
 
\begin{figure*}[ht]
\centering
\includegraphics[width=6cm]{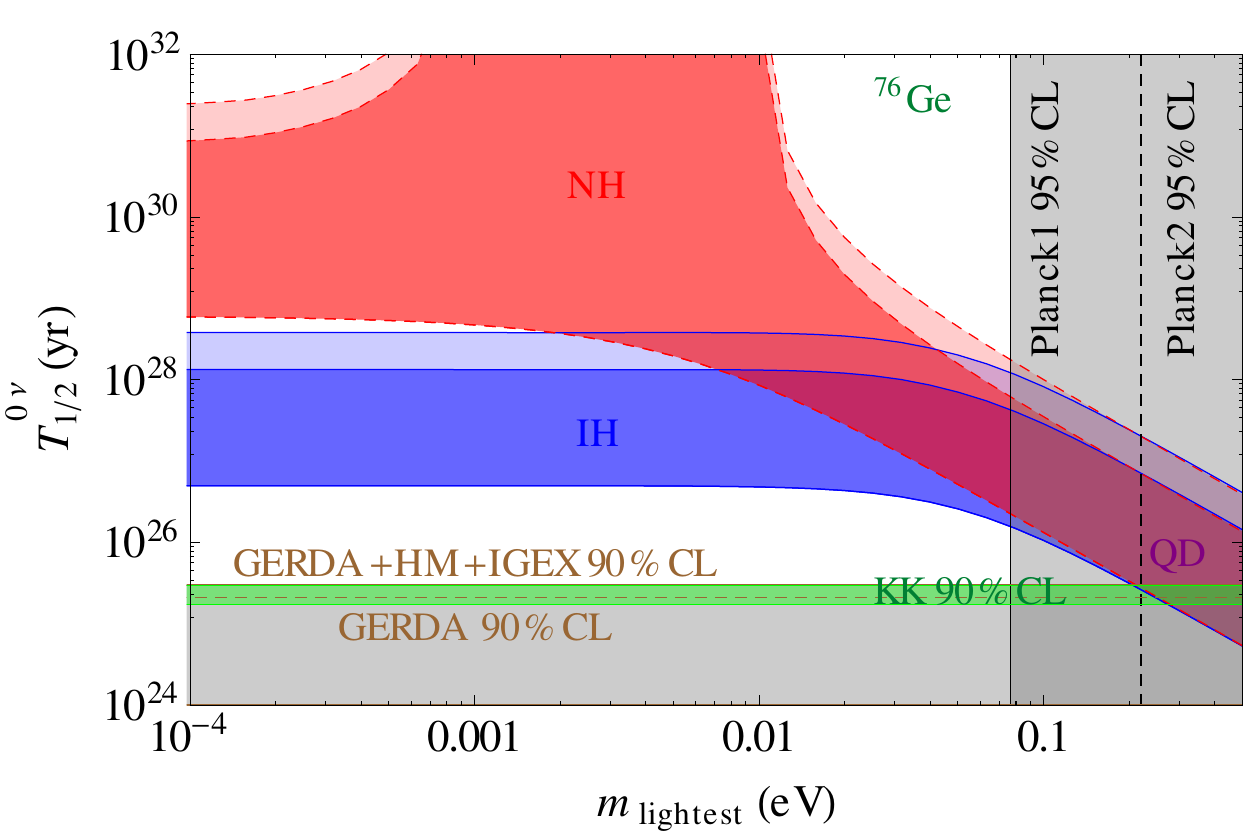}
\hspace{0.5cm}
\includegraphics[width=6cm]{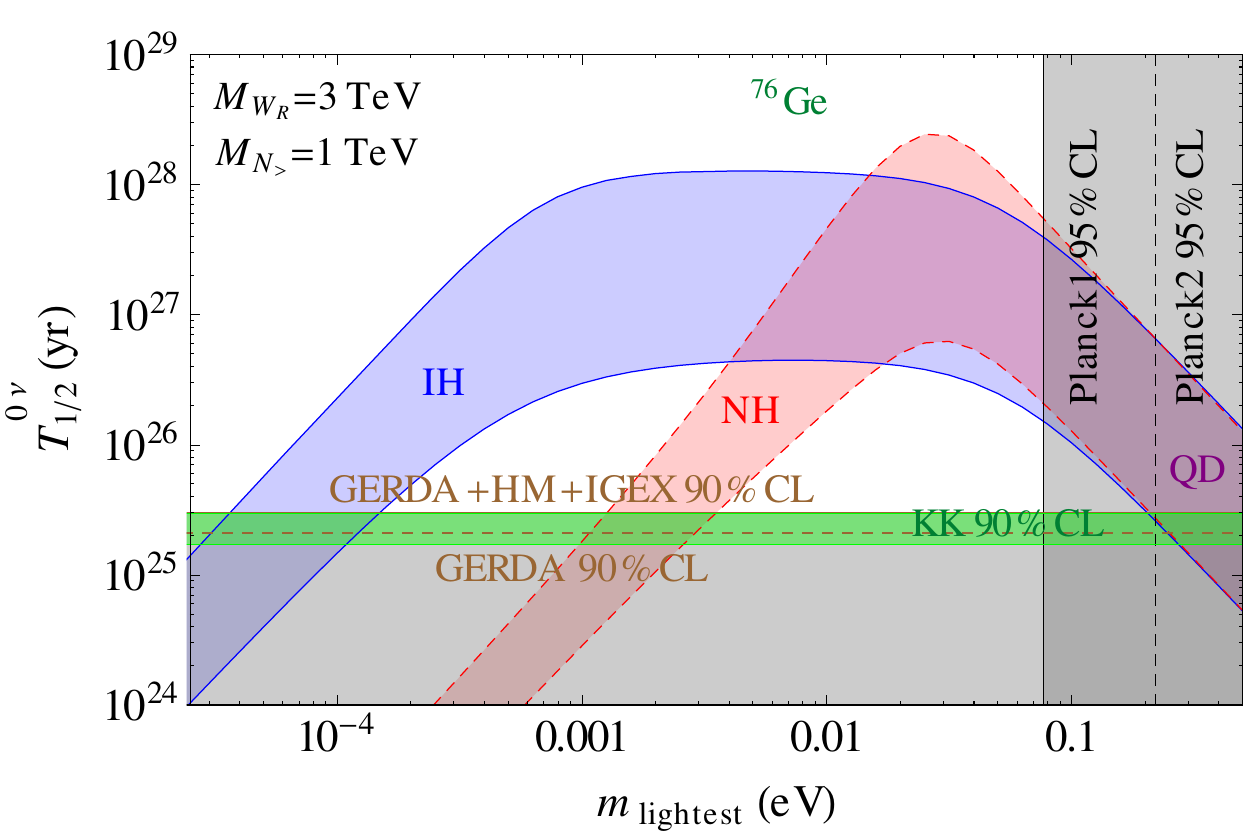}
\caption{The updated Fig.~\ref{fig1} (left panel) and Fig.~\ref{fig2} (right panel) for $^{76}$Ge after including the recent GERDA phase I results.
}
\label{fig5}
\end{figure*}

{\textit{\textbf{Note Added}}}-- After submission of our paper, new results were announced 
from phase I of GERDA experiment~\cite{gerda}, which set new limits on $0\nu\beta\beta$ half-life of $^{76}$Ge: $T_{1/2}^{0\nu}(^{76}{\rm Ge})>2.1\times 10^{25}$ yr at 90\% CL, and when combined with other Ge-based experiments, namely, HM~\cite{hdlbrg} and IGEX~\cite{igex}, it becomes $T_{1/2}^{0\nu}(^{76}{\rm Ge})>3.0\times 10^{25}$ yr at 90\% CL. This new result disfavors the KK claim~\cite{Klapdor:2006ff} independent of NME and of the physical mechanism for $0\nu\beta\beta$. In view of these new results, we show the updated Figs.~\ref{fig1} and \ref{fig2} for $^{76}$Ge in Fig.~\ref{fig5}. Our conclusion remains unchanged that the canonical light neutrino contribution by itself cannot saturate the GERDA+HM+IGEX 
limit, irrespective of the NME uncertainties. After taking into account the new contributions from a TeV-scale LR model with Type-II seesaw,  this limit can be saturated. However it puts a lower  limit on the lightest
neutrino mass in the range of (2-4) meV for NH and (0.03-0.2) meV for IH in the context of this model.  

For completeness, we also compare the corresponding upper limits on the effective neutrino mass using the recent results for $^{76}$Ge and $^{136}$Xe. For all the NMEs given in Table~\ref{tab1}, our results for the effective neutrino mass due to canonical light neutrino contribution are given in Table~\ref{tab2}. For comparison, we also give the corresponding ranges preferred by the claimed observation in~\cite{Klapdor:2006ff}.  It is again clear, as in Table~\ref{tab1}, that 
the KK claim, while compatible with the individual limits from KLZ for some of 
the QRPA NMEs in~\cite{Meroni:2012qf, 
Simkovic:2013qiy, Mustonen:2013zu}, is inconsistent with the combined (KLZ+EXO) limit for all NMEs, except~\cite{Mustonen:2013zu}.  Also we find that the limits on $m_{ee}^\nu$ derived from $^{136}$Xe  are stronger than those from $^{76}$Ge for all the NMEs, except~\cite{Mustonen:2013zu} in which case the two limits are similar.   
\begin{table}[h!]
\begin{center}
\begin{tabular}{c|c|c|c|c|c}\hline\hline
& \multicolumn {5}{|c}{Limit on $m^{\nu}_{ee}$ (eV)}\\ \cline{2-6} 
NME & \multicolumn{3}{|c}{$^{76}$Ge} & \multicolumn{2}{|c}{$^{136}$Xe} \\ \cline{2-6}
& GERDA & comb & KK & KLZ & comb \\ \hline
EDF(U)~\cite{Rodriguez:2010mn} & 0.32 & 0.27 & 0.27-0.35 & 0.15 & 0.11 \\ 
ISM(U)~\cite{Menendez:2008jp} &  0.52 & 0.44 & 0.44-0.58 & 0.28 & 0.21 \\ 
IBM-2~\cite{Barea:2013bz} &  0.27 & 0.23 & 0.23-0.30 & 0.19 & 0.14 \\ 
pnQRPA(U)~\cite{Suhonen:2010zzc} &  0.28 & 0.24 & 0.24-0.31 & 0.20 & 0.15 \\
SRQRPA-B~\cite{Meroni:2012qf} & 0.25 & 0.21 & 0.21-0.28 & 0.18 & 0.14 \\ 
SRQRPA-A~\cite{Meroni:2012qf} & 0.31 & 0.26 & 0.26-0.34 & 0.27 & 0.20 \\ 
QRPA-B~\cite{Simkovic:2013qiy} & 0.26 & 0.22 & 0.22-0.29 & 0.25 & 0.19 \\ 
QRPA-A~\cite{Simkovic:2013qiy} & 0.28 & 0.24 & 0.24-0.31 & 0.29 & 0.21 \\ 
SkM-HFB-QRPA~\cite{Mustonen:2013zu} &  0.29 & 0.24 & 0.24-0.32 & 0.33 & 0.25 \\ 
\hline\hline
\end{tabular}
\end{center}
\caption{The upper limits on the effective neutrino mass $m^\nu_{ee}$ corresponding to the 90\% CL lower bounds on half-lives of $^{76}$Ge (from GERDA and GERDA+HM+IGEX combined~\cite{gerda}) and $^{136}$Xe (from KLZ and KLZ+EXO combined~\cite{Gando:2012zm}) for different NME calculations~\cite{Rodriguez:2010mn, Menendez:2008jp, Barea:2013bz, Suhonen:2010zzc, Meroni:2012qf, Simkovic:2013qiy, Mustonen:2013zu}. Also shown are its preferred ranges corresponding to the 90\% CL half-life of $^{76}$Ge from the KK claim~\cite{Klapdor:2006ff}.  }
\label{tab2}
\end{table}

Similarly, for the heavy neutrino contribution in our minimal LR model, 
we can derive an upper limit on the quantity $M_{W_R}^{-4}\sum_j V^2_{ej}/M_j$ given in Eq.~(\ref{mNee}) using the experimental lower limits on $T_{1/2}^{0\nu}$. Our results are given in Table~\ref{tab3} for the NMEs 
in~\cite{Meroni:2012qf}. Here ``Argonne" and ``CD-Bonn" stand for different nucleon-nucleon potentials, and ``large" or ``intm" refer to different size of the single-particle spaces in the model.  From Table~\ref{tab3} we see that even with the heavy neutrino contribution, the incompatibility between the KK claim and the recent combined limits from $^{76}$Ge and $^{136}$Xe experiments 
still persists. Moreover, the limits from $^{136}$Xe on the parameter characterizing the heavy neutrino contribution to $0\nu2\beta$ are found to be stronger than those from $^{76}$Ge.   
\begin{table}[h!]
\begin{center}
\begin{tabular}{c|c|c|c|c|c}\hline\hline
SRQRPA & \multicolumn {5}{|c}{Limit on $M_{W_R}^{-4}\sum_j V^2_{ej}/M_j~({\rm TeV}^{-5})$}\\ \cline{2-6} 
NME & \multicolumn{3}{|c}{$^{76}$Ge} & \multicolumn{2}{|c}{$^{136}$Xe} \\ \cline{2-6}
method & GERDA & comb & KK & KLZ & comb \\ \hline
Argonne intm & 0.30 & 0.24 & 0.24-0.33 & 0.18 & 0.13 \\ 
Argonne large & 0.26 & 0.22 & 0.22-0.29 & 0.18 & 0.14 \\  
CD-Bonn intm & 0.20 & 0.16 & 0.17-0.22 & 0.17 & 0.13 \\ 
CD-Bonn large & 0.17  & 0.14 & 0.14-0.18 & 0.17 & 0.13\\  
\hline\hline
\end{tabular}
\end{center}
\caption{The upper limits on the heavy neutrino effective mass parameter  $M_{W_R}^{-4}\sum_j V^2_{ej}/M_j$ corresponding to the 90\% CL lower bounds on half-lives of $^{76}$Ge (from GERDA and GERDA+HM+IGEX combined~\cite{gerda}) and $^{136}$Xe (from KLZ and KLZ+EXO combined~\cite{Gando:2012zm}) for the heavy neutrino NMEs in~\cite{Meroni:2012qf}. Also shown are its preferred ranges corresponding to the 90\% CL half-life of $^{76}$Ge from the KK claim~\cite{Klapdor:2006ff}.  }
\label{tab3}
\end{table}

\end{document}